# Interlayer electronic coherence links magnetism and superconductivity in Ruddlesden–Popper nickelates

Feiyang Liu[1], Lixing Chen[1], Enkang Zhang[1], Ying-Jie Zhang[1], Jun Zhao[1,2,3]*

[1]*State Key Laboratory of Surface Physics and Department of Physics, Fudan University, Shanghai 200433, China*

[2]*Shanghai Research Center for Quantum Sciences, Shanghai 201315, China*

[3]*Institute of Nanoelectronics and Quantum Computing, Fudan University, Shanghai 200433, China*

Abstract

The extent to which electronic dimensionality influences magnetism and superconductivity in Ruddlesden–Popper (RP) nickelates remains unsettled. Here we report high-precision crystallographic-axis–resolved dc transport measurements on high-quality single crystals of bilayer and trilayer RP nickelates. Using a six-terminal geometry, we self-consistently determine the intrinsic in-plane ($\rho_{\parallel}$) and out-of-plane ($\rho_{\perp}$) resistivities on the same crystal, while minimizing uncertainties associated with current redistribution in highly anisotropic conductors. We uncover strong intrinsic electronic anisotropy in both bilayer and trilayer nickelates, in contrast to the weak anisotropy inferred from conventional four-probe measurements. Moreover, $\rho_{\perp}$ exhibits a nonmonotonic temperature dependence, revealing a universal coherent-to-incoherent crossover in interlayer transport. Across the RP nickelate series, the maximum superconducting transition temperature ($T_c$) observed under pressure is inversely correlated with the ambient-pressure resistivity anisotropy, suggesting that stronger interlayer electronic coherence is favorable for superconductivity. In addition, $\rho_{\perp}$ serves as an exceptionally sensitive and selective probe of magnetic and density-wave orders, exhibiting pronounced anomalies, whereas only weak signatures are observed in $\rho_{\parallel}$. Our results highlight interlayer coherence as a key organizing parameter that both tracks the relevant magnetic correlations and is closely tied to superconductivity, providing stringent constraints on microscopic theories of high-$T_c$ superconductivity in nickelates.

Electronic dimensionality plays a central role in correlated materials, strongly affecting quantum fluctuations, electronic correlations, and the emergence of intertwined phases such as charge- and spin-density waves (DWs), and high-temperature (high-$T_c$) superconductivity. In particular, reduced dimensionality is often considered favorable for high-$T_c$ superconductivity in layered materials such as cuprates and iron-based superconductors [1–6]. The recent discovery of unconventional superconductivity in thin-film infinite-layer nickelates and both pressurized bulk and thin-film Ruddlesden–Popper (RP) nickelates [7–22] represents a significant breakthrough in the quest for new high-$T_c$ superconductors and naturally motivates comparison with the extensively studied cuprates. Structurally, RP nickelates feature Ni-O layers separated by rare-earth-based spacer layers, closely paralleling the layered architecture central to cuprate physics. In cuprates, the quasi-two-dimensional Fermi surface [23,24] derived from the Cu $3d_{x^2-y^2}$ orbital leads to pronounced transport anisotropy, with out-of-plane conductivity orders of magnitude lower than in-plane conductivity [25–29].

Electronic structure calculations, angle-resolved photoemission spectroscopy (ARPES), and optical studies suggest that RP nickelates share a similar quasi-two-dimensional electronic structure and charge dynamics with cuprates [30–38], though with multi-orbital (Ni $3d_{z^2}$ and $3d_{x^2-y^2}$) contributions near the Fermi level [39–43]. This points to broad similarities in low-dimensional electronic behavior between the two systems. Interestingly, however, trilayer RP nickelates exhibit a lower $T_c$ than their bilayer counterparts [8–18,22], in stark contrast to cuprates, where trilayer systems typically host the highest $T_c$ [44–46]. This contrast suggests that electronic dimensionality and interlayer coherence may play qualitatively different roles in RP nickelates.

Yet direct experimental evidence from transport measurements has remained elusive. Previous dc transport measurements using the conventional four-probe method on RP nickelates reported a rather weak resistivity anisotropy, $\rho_{\perp}/\rho_{\parallel} < 10$ at ambient pressure and room temperature [47], which seemingly inconsistent with expectations for a quasi-two-dimensional metal. Interpreting anisotropy in strongly layered conductors, however, is experimentally challenging, because current distribution and contact geometry in standard four-probe configurations can distort the apparent voltage response, particularly when the intrinsic anisotropy is large. This motivates direct, high-precision determination of the in-plane and out-of-plane resistivities, as well as a systematic comparison across the RP nickelate series.

In this paper, we address this issue by performing self-consistent in-plane and out-of-plane dc transport measurements on single crystals of bilayer and trilayer RP nickelates ($La_4Ni_3O_{10}$, $Pr_4Ni_3O_{10}$, and $La_3Ni_2O_7$). High-quality single crystals were grown using the high-pressure optical floating zone technique, as detailed elsewhere [11,15,48]. Laue-oriented crystals were cut into bar-shaped specimens with approximate dimensions of length ($L$) ~ 0.8–1.1 mm, width ($b$) ~ 0.3–0.5 mm and thickness ($D$) ~ 40–60 μm (see Fig. 1e). Subsequently, all faces of the bar were mechanically polished to achieve optically flat surfaces, ensuring reliable contact fabrication. Electrical contacts were fabricated by sputtering gold stripes onto the polished surfaces, followed by attaching 25-μm-diameter gold wires using silver paint. To accurately resolve the strong transport anisotropy and minimize systematic uncertainties associated with contact resistance and current inhomogeneity, we employed a six-terminal measurement geometry [49–53], in which four evenly spaced electrical contacts were patterned on both the top and bottom surfaces along the length of the bar (see Fig. 1d). This approach allows simultaneous and self-consistent determination of in-plane and out-of-plane resistivities on the same specimen, minimizing geometric and contact-related artifacts that are particularly severe for highly anisotropic layered materials [49].

In the six-terminal configuration, current $I$ is applied through two outer contacts on the top surface. Voltages are measured simultaneously across two inner contacts on the top surface ($V_{\mathrm{top}}$) and across the corresponding contacts directly beneath on the bottom surface ($V_{\mathrm{bot}}$). The intrinsic in-plane resistivity $\rho_{\parallel}$ and out-of-plane resistivity $\rho_{\perp}$ are related by the following rapidly converging series [50,51]:

$$V_{\text{bot}} = \frac{8I(\rho_\perp \rho_\parallel)^{1/2}}{\pi b} \sum_{k=0}^{\infty} \frac{(-1)^k}{2k+1} \frac{\sin[(2k+1)\pi l/2L]}{\sinh[(2k+1)\eta]} \tag{1}$$

$$V_{\text{top}} = \frac{8I(\rho_\perp \rho_\parallel)^{1/2}}{\pi b} \left[ \begin{array}{c} \frac{1}{2} \ln \tan\left(\frac{\pi}{4} + \frac{\pi l}{4L}\right) \\ + \sum_{k=0}^{\infty} \frac{(-1)^k}{2k+1} \sin\frac{(2k+1)\pi l}{2L} \\ \times [\coth((2k+1)\eta) - 1] \end{array} \right] \tag{2}$$

Where $\eta \equiv (\rho_\perp/\rho_\parallel)^{1/2}\pi D/L$; $L$, $b$ and $D$ denote the length, width and thickness of the sample, respectively, and $l = (x_1 - x_2)$ represents the distance between the voltage contacts, see Fig. 1d.

Fig. 1a-c presents the effective resistivities of single-crystalline $La_4Ni_3O_{10}$, $Pr_4Ni_3O_{10}$, and $La_3Ni_2O_7$, obtained from voltage signals on top and bottom surfaces, plotted under the conventional assumption of uniform current density across the sample cross-section. A striking feature of the raw data is the large disparity between the voltages measured on opposite sides of our ~50-$\mu$m-thick crystals. In this representation, $\rho_{\text{top}}$ exhibits metallic behaviors with pronounced anomalies near the DW transition temperatures, in agreement with prior reports [30,48,54–56]. In contrast, $\rho_{\text{bot}}$ falls more rapidly upon cooling and approaches zero at low temperatures.

A key implication from equations (1) and (2) is that for a large resistivity anisotropy ($\rho_\perp/\rho_\parallel \gg 1$), the applied current is confined to a thin layer near the top surface, with an effective penetration depth $z_{\text{eff}} \approx L\pi^{-1}(\rho_\parallel/\rho_\perp)^{1/2}$. Consequently, $V_{\text{top}}$ measures the geometric mean of in-plane and out-of-plane resistivity $(\rho_\perp\rho_\parallel)^{1/2}$ rather than the in-plane resistivity $\rho_\parallel$ alone as assumed in the standard treatment. The large disparity between $V_{\text{top}}$ and $V_{\text{bot}}$ measured in our bar-shaped thin crystals confirmed the highly non-uniform current within the sample volume—a direct signature of substantial electrical anisotropy in these layered nickelates. Therefore, the intrinsic, temperature-dependent in-plane and out-of-plane resistivities must be extracted using equations (1) and (2), rather than relying on the uniform-current approximation.

For $La_4Ni_3O_{10}$ and $Pr_4Ni_3O_{10}$ below their intertwined charge- and spin-DW (CDW/SDW) transition temperatures [54,55], the resistivity anisotropy becomes so large that the effective current penetration depth, estimated to be less than 10 $\mu$m for our trilayer nickelate samples, is much smaller than the sample thickness. Under such conditions, the voltage $V_{\mathrm{bot}}$ falls below the resolution of our nanovoltmeter. To overcome this, we adopted an alternative configuration in which $V_{\perp}$ was measured by applying current through opposing contacts on the top and bottom surfaces. Subsequently, $\rho_{\parallel}$ and $\rho_{\perp}$ were obtained by numerically solving equations using the measured values of $V_{\perp}$ and $V_{\mathrm{top}}$ (see Supplemental Material for calculation details).

The resulting intrinsic in-plane resistivities $\rho_{\parallel}$ for all three compounds display qualitatively similar metallic behavior (Fig. 2a, c, e) and exhibit a Fermi-liquid $T^2$ dependence at low temperatures (Fig. 3). Upon warming, anomalies appear near the DW transition temperatures of around 138 K, 158 K and 110 K for $La_4Ni_3O_{10}$, $Pr_4Ni_3O_{10}$, and $La_3Ni_2O_7$, respectively [48,54–63]. Compared with results obtained from the uniform-current assumption, the DW-related anomalies in $\rho_{\parallel}$ are significantly suppressed, highlighting the importance of properly accounting for current redistribution in highly anisotropic systems.

In contrast, the out-of-plane resistivity $\rho_{\perp}$ exhibits an unconventional temperature dependence (Fig. 2b, d, f). At high temperatures, $\rho_{\perp}$ shows apparent non-metallic behavior, decreasing with increasing temperature —a feature common to all three compounds. Upon cooling, $\rho_{\perp}$ of the trilayer compounds increases sharply below the coupled CDW/SDW transitions. In bilayer $La_3Ni_2O_7$, $\rho_{\perp}$ displays two distinct anomalies at 150 K and 110 K, corresponding to SDW and CDW transitions, respectively [64]; notably, the SDW-related feature near 150 K is essentially absent in $\rho_{\parallel}$. In trilayer $Pr_4Ni_3O_{10}$, an additional kink anomaly is observed in $\rho_{\perp}$ at 26 K, and warming and cooling curves bifurcate between this temperature and the DW transition (see Fig. 2d). This behavior coincides with the onset of a crossover from two-dimensional to three-dimensional magnetic order upon cooling, a metastable hysteretic imprinting effect involving the Ni sublattice upon warming, and the development of Pr-related spin correlations [55]. This feature is likewise not discernible in $\rho_{\parallel}$. These results suggest that $\rho_{\perp}$ is strongly coupled to SDW and CDW, whereas $\rho_{\parallel}$ is only weakly responsive to the CDW. Upon further cooling, $\rho_\perp$ passes through a broad maximum. At sufficiently low temperatures, the $\rho_\perp$ of all three compounds recovers a Fermi-liquid $T^2$ dependence (see Fig. 3).

The nonmonotonic temperature dependence of the out-of-plane resistivity can be understood as a crossover from a low-temperature regime with coherent interlayer (band-like) transport to a higher-temperature regime in which single-particle $k_z$ coherence is progressively lost. At low temperatures, the recovery of a Fermi-liquid-like $T^2$ dependence in $\rho_\perp$ is consistent with finite interlayer electronic coherence. As temperature increases, once the scattering rate becomes comparable to the effective interlayer hopping scale, scattering broadening washes out the interlayer dispersion and progressively destroys single-particle coherence along $k_z$, naturally accounting for the crossover in $\rho_\perp(T)$. The distinct temperature dependences of $\rho_\parallel$ and $\rho_\perp$ indicate that the out-of-plane transport cannot be understood from a single isotropic scattering rate of quasiparticles alone, but must also involve the evolution of interlayer coherence and correlation effects [65–68].

The much more pronounced anomalies in $\rho_\perp(T)$ at DW and magnetic transitions, compared with the relatively weak signatures in $\rho_\parallel(T)$, suggest that these ordered states couple disproportionately to the electronic degrees of freedom governing interlayer charge transport. A natural interpretation is that the ordering either partially gaps, or strongly renormalizes, the $k_z$-dispersive electronic states that carry most of the out-of-plane current, thereby producing an amplified response in $\rho_\perp$ [30,37,38,69,70]. Overall, these results point to an intrinsic linkage between interlayer electronic coherence and magnetic or DW correlations in RP nickelates.

Having determined the intrinsic resistivities along both crystallographic directions, we extract the temperature-dependent resistivity anisotropy ratio, $\gamma_\rho \equiv \rho_\perp/\rho_\parallel$, for all three compounds, as shown in Fig. 4a. While earlier studies on trilayer nickelates reported rather weak anisotropy [47,48], our measurements reveal a much larger anisotropy, with $\gamma_\rho$ saturating at values of 1.8 × $10^4$, 1.8 × $10^3$, and 1.7 × $10^2$ in the low-temperature Fermi-liquid regimes for $La_4Ni_3O_{10}$, $Pr_4Ni_3O_{10}$, and $La_3Ni_2O_7$, respectively. This aligns with the expectations from the quasi-two-dimensional Fermi surfaces arising from Ni 3*d* orbitals in these materials [34]. Upon warming, $\gamma_\rho$ decreases monotonically and exhibits an additional sharp drop through the DW transitions in the trilayer compounds but $\gamma_\rho$ remains substantial even at room temperature, highlighting the persistence of strongly anisotropic transport across a broad temperature range.

To assess whether lattice changes can account for the anisotropy enhancement below the DW transitions, we measured the out-of-plane thermal expansion of $La_4Ni_3O_{10}$, which exhibits the largest $\gamma_\rho$ in this family, using a high-resolution capacitive dilatometer [71] (see Fig. 4b). Upon cooling across the DW transition, the crystal contracts along the out-of-plane direction, yet $\gamma_\rho$ increases sharply. This opposite trend rules out a simple geometric explanation based on interlayer spacing and instead points to an electronic mechanism tied to DW–driven magnetic reconstruction.

Comparing compounds, $La_4Ni_3O_{10}$ is more anisotropic than $Pr_4Ni_3O_{10}$, and the bilayer compound $La_3Ni_2O_7$ displays the smallest resistivity anisotropy in the low-temperature Fermi-liquid regime. This trend runs counter to density-functional-theory expectations that the electronic structure of layered RP nickelates $R_{n+1}Ni_nO_{3n+1}$ becomes more three-dimensional with increasing layer number *n* [34]. This apparent reversal suggests that interlayer electronic coherence is not governed solely by geometric layering, but is instead strongly influenced by orbital character and magnetic correlations. In particular, the enhanced interlayer transport in $La_3Ni_2O_7$ is consistent with a more effective contribution from $d_{z^2}$-derived states to interlayer hopping, whereas in trilayer compounds, the presence of an inner Ni-O layer with suppressed magnetism may reduce coherent interlayer coupling. These results indicate that bilayer RP nickelates can sustain more coherent interlayer transport than their trilayer counterparts, despite their reduced structural dimensionality.

A central question is whether interlayer coherence also bears on superconductivity under pressure. In Fig. 4c, we compare low-temperature $\gamma_\rho$ with the maximum superconducting transition temperatures reported under pressure [8,11,15] and a clear trend is revealed: the transport anisotropy anticorrelates with $T_c$ across these RP nickelates. This suggests that enhanced interlayer electronic coherence is favorable for superconductivity in this family. Although pressure modifies several microscopic parameters simultaneously, this empirical trend provides a useful perspective on why the layer-number dependence of $T_c$ in RP nickelates differs from that in multilayer cuprates, and highlights a distinct role of interlayer coupling in the nickelate setting.

It is also interesting to view this trend in a broader nickelate context across different structural families. Beyond RP nickelates, an inverse correlation between the $c$-axis lattice constant and $T_c$ has been recently reported in infinite-layer nickelate thin films [72,73], suggesting that out-of-plane structural control may likewise influence superconductivity. This comparison raises an intriguing possibility that interlayer coherence acts as a "hidden control knob" for nickelate superconductivity, even though the structural ingredients that enable it differ significantly between families. In the RP compounds, tuning the apical-oxygen-assisted involvement of the Ni $3d_{z^2}$ orbitals can directly reshape the $k_z$ dispersion and the coupling between charge/spin correlations across layers, making the transport anisotropy a particularly sensitive diagnostic. By contrast, in infinite-layer nickelates, where the apical oxygen channel is absent and electronic structure is more strongly dominated by $d_{x^2-y^2}$ character, analogous trends could arise from $c$-axis chemical pressure, interface-enabled orbital reconstruction, and rare-earth band hybridization, each of which may increase three-dimensional coherence without relying on the same orbital pathway. If so, the inverse $c$-axis–$T_c$ trend in infinite-layer films and the $T_c$–anisotropy trend in RP nickelates may represent distinct microscopic pathways converging on a common macroscopic outcome. Such a cross-family correspondence suggests a unifying organizing principle for correlated nickelates and sharply narrows the space of viable microscopic pairing scenarios.

In summary, we performed crystallographic-axis–resolved dc transport measurements on high-quality single crystals of bilayer and trilayer RP nickelates using a six-terminal geometry that self-consistently determines the intrinsic in-plane and out-of-plane resistivities on the same specimen, uncovering a pronounced electronic anisotropy. The out-of-plane resistivity $\rho_{\perp}$ exhibits nonmonotonic temperature dependence in both bilayer and trilayer compounds, revealing a universal coherent-to-incoherent crossover in interlayer transport. The resistivity anisotropy is found to anticorrelate with the maximum superconducting transition temperature achieved under pressure, indicating that enhanced interlayer electronic coherence is favorable for superconductivity in this family. Moreover, the pronounced response of interlayer transport to SDW orders, in contrast to the comparatively weak in-plane signatures, suggests a strong coupling between magnetism and the interlayer conduction channel dominated by Ni $3d_{z^2}$ orbitals. These results establish interlayer electronic coherence as a key organizing principle linking magnetism, dimensionality, and superconductivity in RP nickelates, and provide essential experimental constraints for microscopic theories of nickelate high-$T_{\mathrm{c}}$ superconductivity.

**Acknowledgements**

This work was supported by the Key Program of the National Natural Science Foundation of China (12234006), the National Key R&D Program of China (2022YFA1403202), the Quantum Science and Technology-National Science and Technology Major Project (2024ZD0300100), the Shanghai Municipal Science and Technology Project (25DZ3008100, 2019SHZDZX01), and the Large Scientific Facility Open Subject of Songshan Lake Laboratory (KFKT2022A03). Part of the sample fabrication was performed at Fudan Nano-fabrication Laboratory.

*Corresponding author: zhaoj@fudan.edu.cn

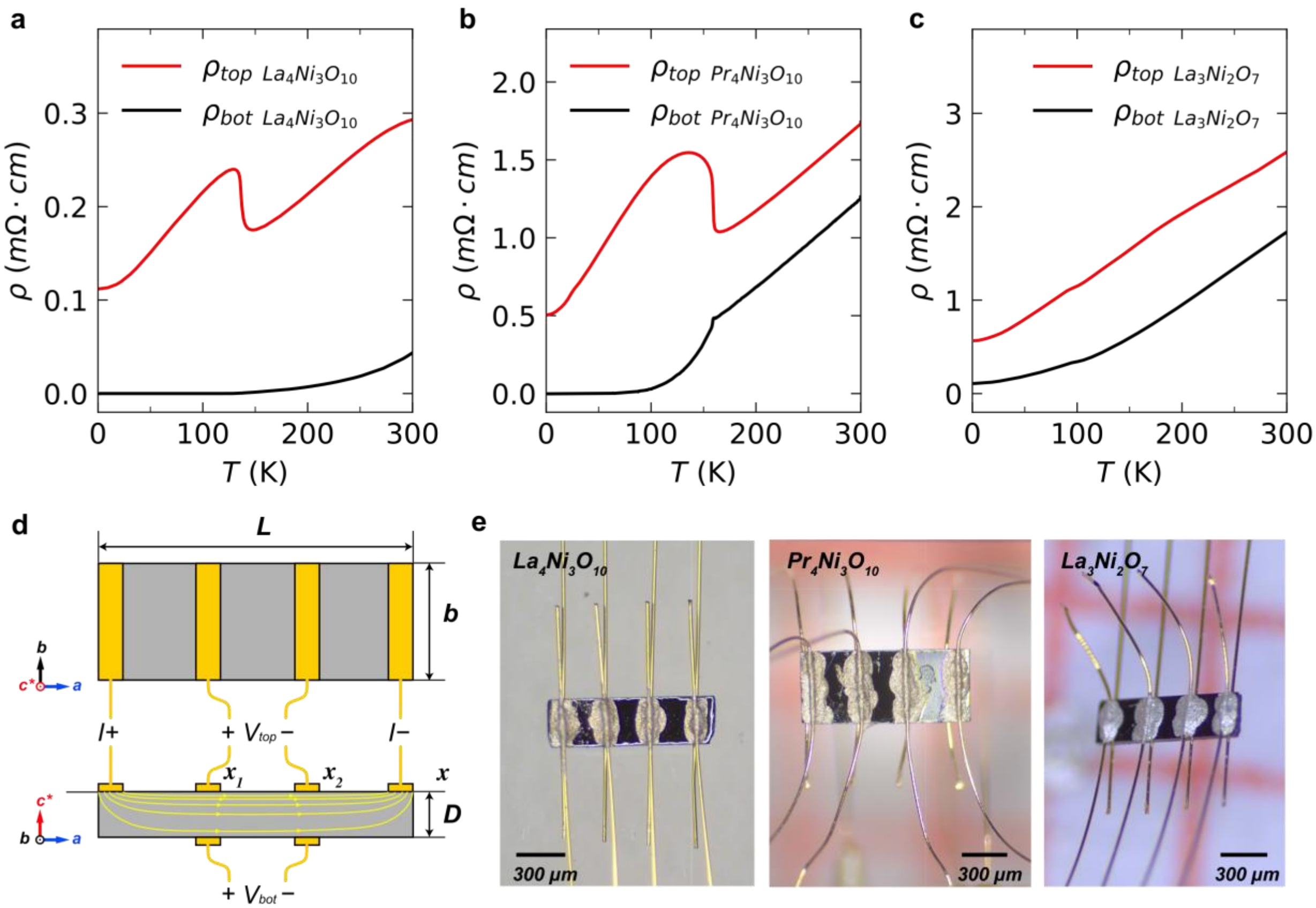

**Figure 1.** Six-terminal resistivity measurements on $La_4Ni_3O_{10}$, $Pr_4Ni_3O_{10}$, and $La_3Ni_2O_7$. **(a–c)** Nominal resistivities $\rho_{\mathrm{top}}$ and $\rho_{\mathrm{bot}}$ of $La_4Ni_3O_{10}$, $Pr_4Ni_3O_{10}$, and $La_3Ni_2O_7$, respectively, calculated from voltages recorded on top and bottom sample surfaces of each sample. **(d)** Schematic illustration of the current distribution in an anisotropic sample measured using the six-terminal method. For highly anisotropic materials, the current density is non-uniform across the sample cross-section. Here, ***c**** denotes the direction normal to the *ab*-plane. **(e)** Images of bar-shaped single crystals measured using the six-terminal configuration.

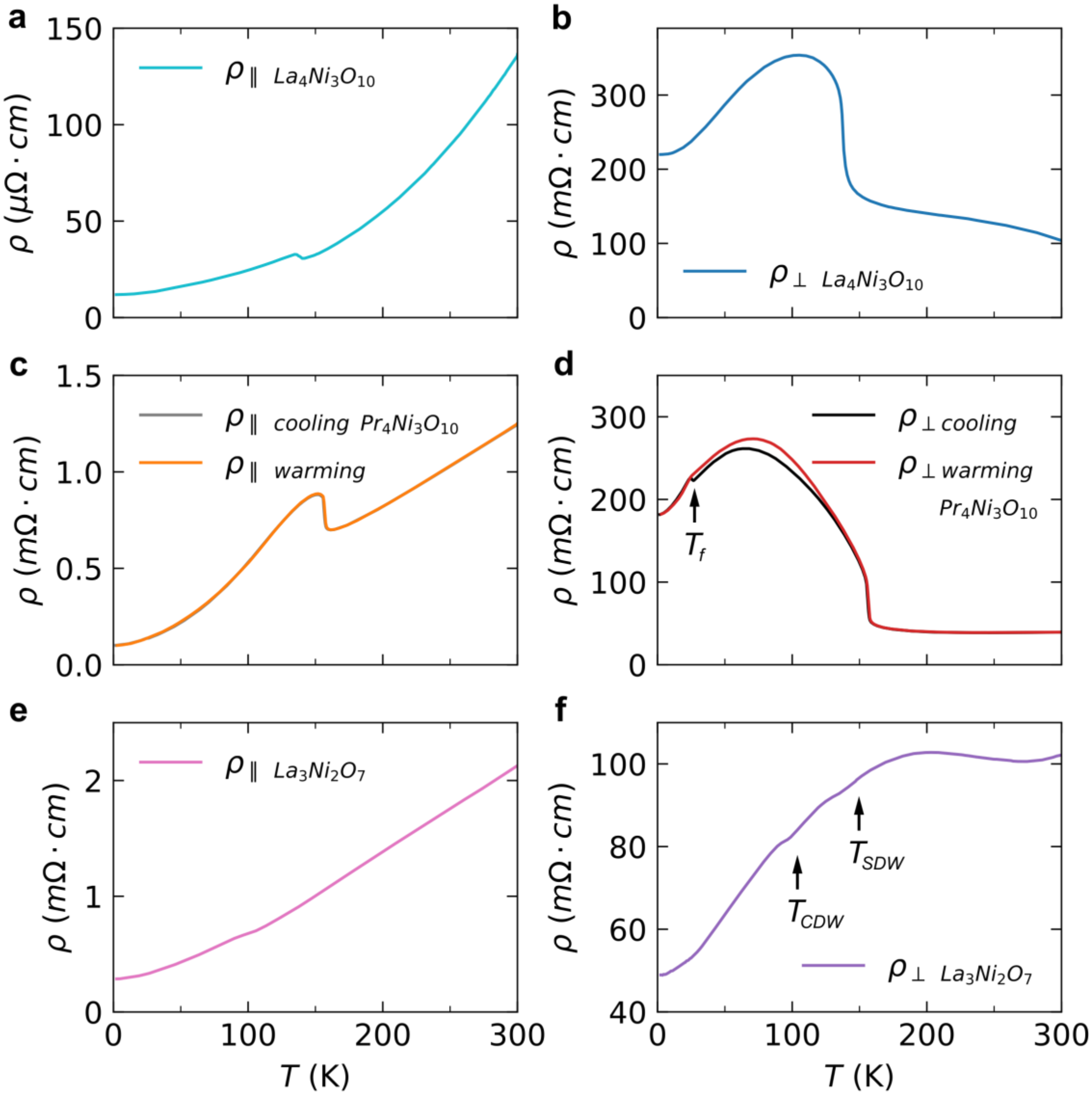

**Figure 2.** Extracted in-plane and out-of-plane resistivities for $La_4Ni_3O_{10}$, $Pr_4Ni_3O_{10}$, and $La_3Ni_2O_7$. **(a, c, e)** In-plane resistivity $\rho_{\parallel}(T)$ of $La_4Ni_3O_{10}$, $Pr_4Ni_3O_{10}$, and $La_3Ni_2O_7$, respectively. For $Pr_4Ni_3O_{10}$, warming and cooling curves are shown in orange and grey, respectively, and are nearly completely overlapped. **(b, d, f)** Out-of-plane resistivity $\rho_{\perp}(T)$ of $La_4Ni_3O_{10}$, $Pr_4Ni_3O_{10}$, and $La_3Ni_2O_7$, respectively. For $Pr_4Ni_3O_{10}$, warming and cooling curves are displayed in red and black; a visible bifurcation between them emerges above the dimensional SDW crossover temperature $T_f$ [55], and they rejoin above the coupled CDW/SDW transition temperature.

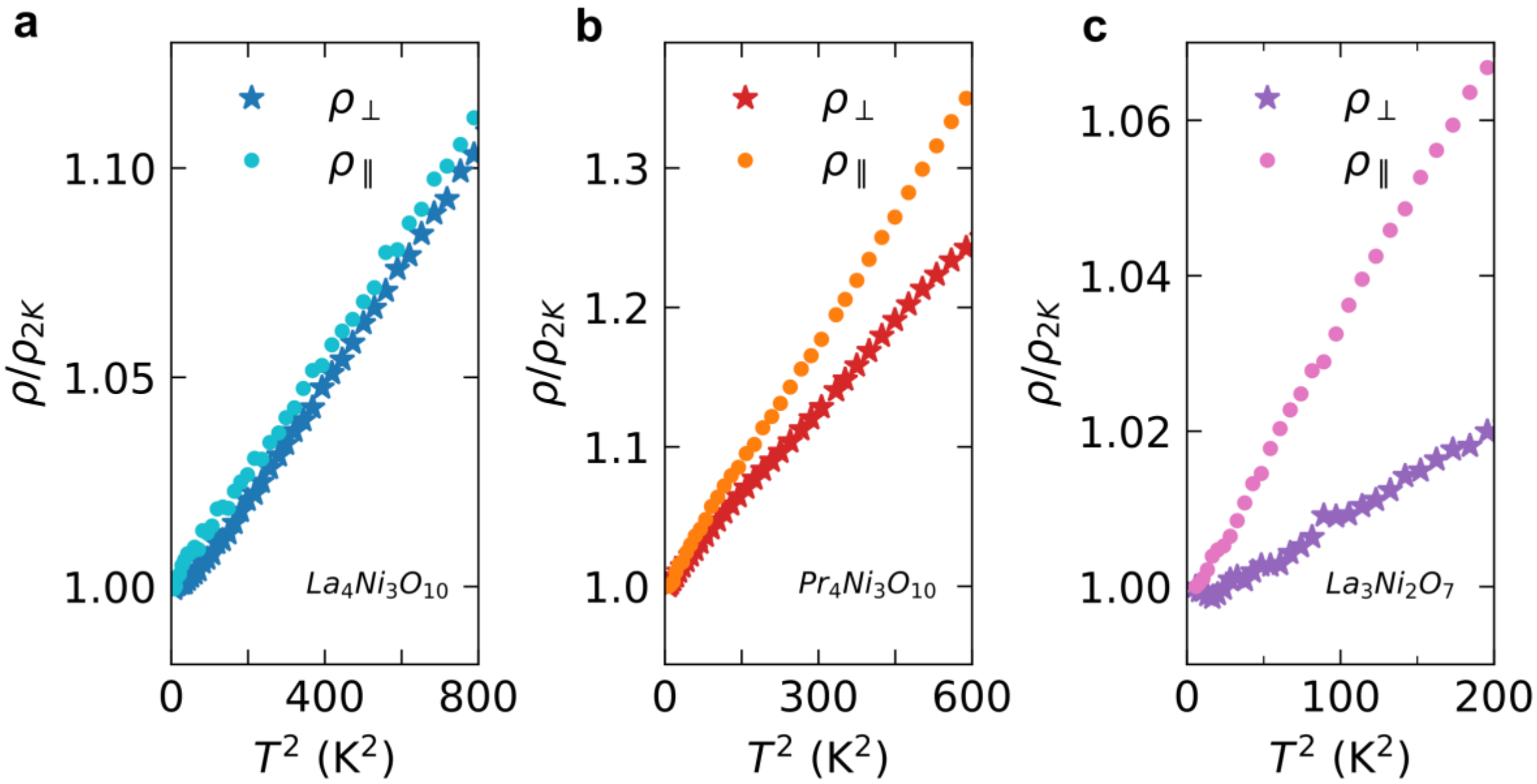

**Figure 3.** Low temperature resistivities of $La_4Ni_3O_{10}$, $Pr_4Ni_3O_{10}$, and $La_3Ni_2O_7$. Both in-plane and out-of-plane resistivities exhibit Fermi-liquid behavior at low temperatures for all three compounds. Data for $Pr_4Ni_3O_{10}$ are taken from warming measurements.

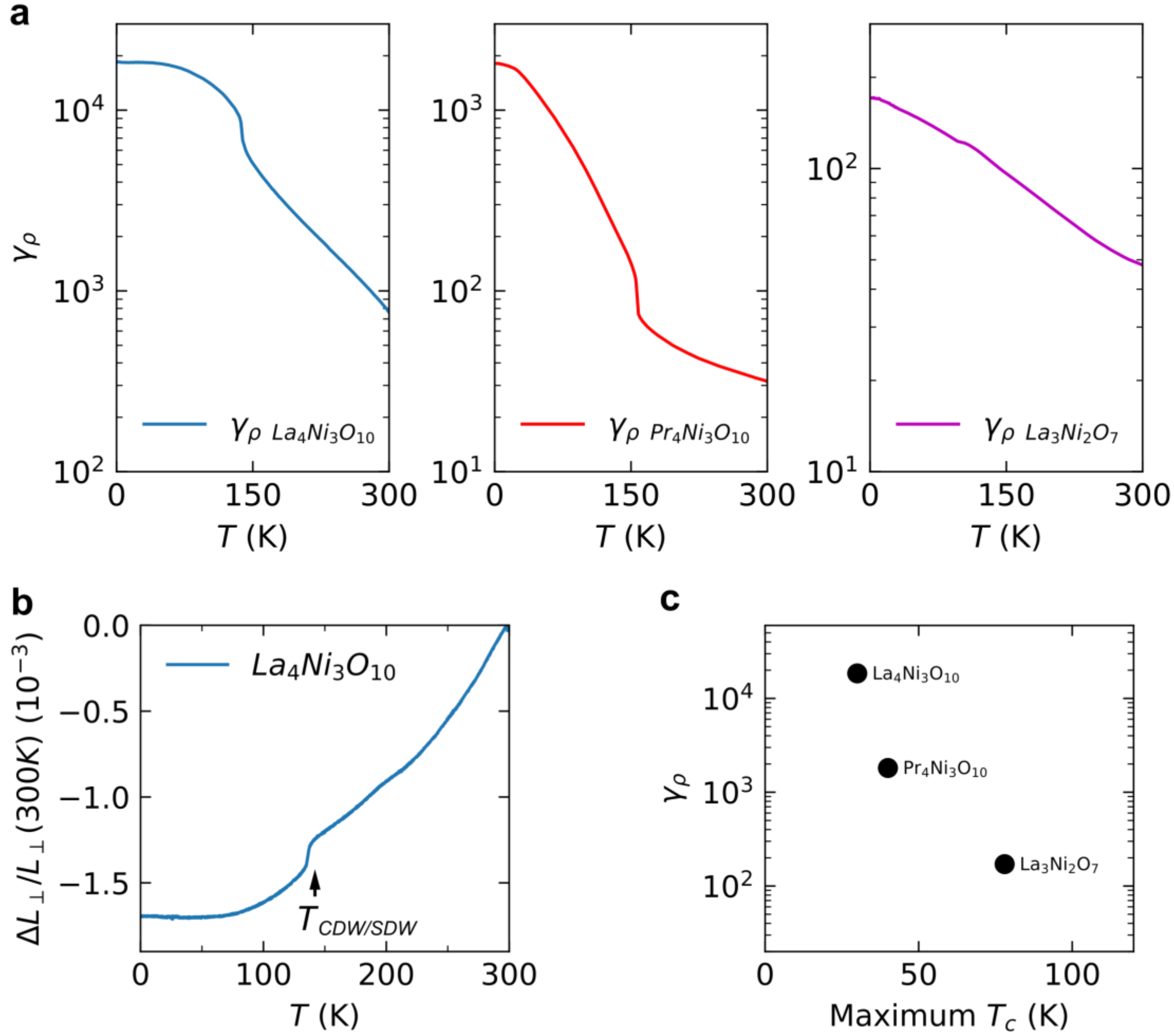

**Figure 4. (a)** Temperature dependence of the resistivity anisotropy $\gamma_\rho \equiv \rho_\perp/\rho_\parallel$ for $La_4Ni_3O_{10}$, $Pr_4Ni_3O_{10}$, and $La_3Ni_2O_7$. Data for $Pr_4Ni_3O_{10}$ are taken from warming measurements. **(b)** Out-of-plane thermal expansion of $La_4Ni_3O_{10}$, showing lattice contraction along the ***c****-direction upon cooling through the concomitant CDW/SDW transition. (**c**) Correlation between the low-temperature (2K) transport anisotropy $\gamma_\rho$ and the maximum superconducting temperature $T_c$ achieved under pressure for the three RP nickelates investigated in this study [8,11,15].